\documentclass[a4paper]{PoS}
\DeclareMathAlphabet{\mathcal}{OMS}{cmsy}{m}{n}

\newcommand{\etal}{{et al.}}
\usepackage{lineno}
\title{Selection of AGN to study the extragalactic background light with HAWC}
\ShortTitle{AGN sample to study the EBL with HAWC}
\author{\speaker{Sara Couti\~no De Le\'on$^{a}$}, Alberto Carrami\~nana Alonso$^{a}$ and Daniel Rosa-Gonz\'alez$^{a}$ for the HAWC Collaboration $^{b}$ \\
\llap{$^{a}$}
Instituto Nacional de Astrof\'isica, \'Optica y Electr\'onica\\
\llap{$^b$}
For a complete author list, see \href{http://www.hawc-observatory.org/collaboration/icrc2015.php}{www.hawc-observatory.org/collaboration/icrc2015.php}.\\
E-mail: \email{sara@inaoep.mx}, \email{alberto@inaoep.mx}, \email{danrosa@inaoep.mx}}

\abstract{The extragalactic background light (EBL) is all the electromagnetic energy released by resolved and unresolved extragalactic sources since the recombination era. Its intensity and spectral shape provide information about the evolution of galaxies throughout cosmic history. Since direct observations of the EBL are very difficult to perform, the study of the interaction between the low energy EBL photons and high energy photons from distant extragalactic sources becomes relevant to constrain the EBL intensity. The main goal of this study is to investigate the opacity of the EBL to gamma rays by observing a sample of active galaxies nuclei (AGN) with the High Altitude Water Cherenkov (HAWC) Gamma-Ray Observatory. Current gamma-ray observations up to 20 TeV performed by Imaging Atmospheric Cherenkov Telescopes (IACTs) have constrained the EBL intensity in the 0.1-50 $\mu$m region. HAWC, which monitors the gamma-ray sky in the 100 GeV to 100 TeV energy range, will be able to detect about 10 AGN with the first year of HAWC data based on the extrapolation of steady-state spectra from the GeV band to TeV, and thus constrain the EBL in the poorly-measured 1-100 $\mu$m range.}
\FullConference{The 34th International Cosmic Ray Conference,\\
		30 July- 6 August, 2015\\
		The Hague, The Netherlands}

\begin{document}

\section{Introduction}
The extragalactic background light (EBL) contains all the electromagnetic radiation released from resolved and unresolved galaxies since the recombination era. The EBL is the second most dominant radiation field in the universe, surpassed by the cosmic microwave background (CMB). Direct measurements of the EBL are difficult at most wavelengths, specially at $\sim15$ $\mu$m where the contribution of the zodiacal light is some orders of magnitude larger than the EBL. Its spectrum is bimodal with one component peaking at $\sim1$ $\mu$m, which corresponds to stellar radiation, and a second component peaking at $\sim100$ $\mu$m, created by the absorption of UV/optical light that is re-radiated by dust at infrared wavelengths. The EBL intensity and spectral shape, therefore, contains important information about the cosmic evolution of these astrophysical objects, the obscuring dust, and the relative contributions of starburst galaxies and AGN to the energy released over cosmic time \cite{2013APh....43..112D}.\\
The High Altitude Water Cherenkov (HAWC) Observatory is a ground-based TeV gamma-ray observatory in the state of Puebla, Mexico at an altitude of 4100 m a.s.l. Completed in March 2015, the detector continuously measures the arrival time and direction of gamma-ray primaries within its 2 sr field of view. It is most sensitive to gamma-ray energies ranging from 100 GeV - 100 TeV. Basic cuts on the data can be used to differentiate gamma-ray air showers from the large cosmic-ray background and employs cuts to differentiate them from cosmic-rays  \cite{hawcsensi,andy,Paco}.\\
The main goal of this work is to select a sample of AGN to study the opacity of the EBL to gamma rays by observing in the 100 GeV to 100 TeV energy range. The spectra observed with HAWC will be corrected using an EBL model in order to estimate the the de-absorbed spectrum. By comparing the de-absorbed fluxes to the intrinsic spectra predicted by AGN models, we can constrain the EBL in the poorly-measured 1-100 $\mu$m region.
\section{Opacity to gamma rays}\label{opacity}
\begin{figure}
\begin{center}
\includegraphics[width=\textwidth , height=11cm]{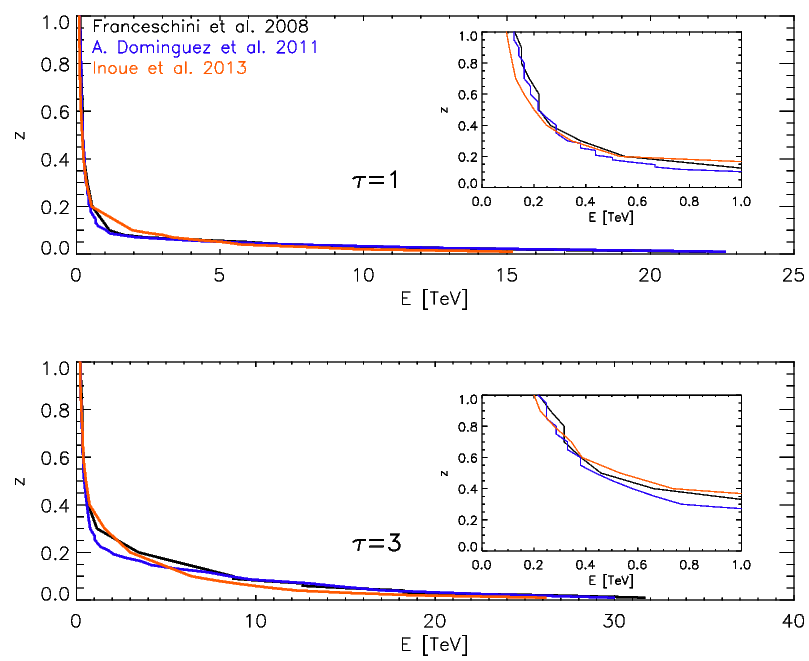}
\caption{Relation between the observed energy and redshift for $\tau=1$ (upper panel) and $\tau=3$ (lower panel) predicted for 3 different EBL models (\cite{2011MNRAS.410.2556D},\cite{2008A&A...487..837F},\cite{2013ApJ...768..197I})}\label{ze}
\end{center}
\end{figure}

On route to Earth, very high-energy gamma rays from cosmological sources have to pass through the radiation field of the EBL, resulting in their attenuation by pair producing interactions with photons in the near- to mid-IR regime of the EBL \cite{1967PhRv..155.1408G}. For a photon emitted by a source located at redshift $z_s$ and observed at $z=0$ with observed energy $E_{\gamma}$, the optical depth between $z_s$ and $z_0$ ($0<z_0<z_s$) is:
\begin{equation}
\tau_{\gamma\gamma}(E_{\gamma},z_0,z_s)=\int_{z_{0}}^{z_{s}}dz\frac{d\ell}{dz}\int_{-1}^{1}d\mu\frac{1-\mu}{2}\int_{\epsilon{}_{th}}^{\infty}\frac{dn_{\epsilon}(\epsilon,z)}{d\epsilon}\sigma_{\gamma\gamma}(E_{\gamma}(1+z),\epsilon,\theta),
\end{equation}
where $\mu=\cos\theta$, $\frac{dn_{\epsilon}}{d\epsilon}$ is the EBL proper photon density, $\frac{d\ell}{dz}=c|\frac{dt}{dz}|$, with $\ell$ the proper distance, $\theta$ is the interaction angle between the two photons, $\sigma_{\gamma\gamma}(E_{\gamma}(1+z),\epsilon,\theta)$ is the cross section of this interaction and $\epsilon_{th}$ is the pair production threshold energy:
\begin{equation}
\epsilon_{th}=\frac{2m_e^2c^4}{E_{\gamma}(1+z)(1-\mu)}.
\end{equation}
The optical depth is a function of the observed energy and the source redshift, therefore the detection of extragalactic sources with HAWC is limited by redshift. If one expects to measure the energy spectrum of a source at $\sim50$ TeV then its redshift must be below 0.1 (Figure \ref{ze}). The observed spectrum will be $\sim37\%$ of intrinsic spectrum for $\tau=1$ (equation \ref{fluxes_atenuation}).
\begin{equation}\label{fluxes_atenuation}
\left(\frac{dN}{dE}\right)_{\tiny{intrinsic}}=\exp[\tau_{\gamma\gamma}(E,z)]\times\left(\frac{dN}{dE}\right)_{\tiny{observed}}.
\end{equation}
The predicted opacity values given by \cite{2008A&A...487..837F} were interpolated in redshift and energy to find the redshift limit for HAWC's observations and it was found that sources with $z<0.5$ are mostly likely to be detected at very high energies with HAWC.\\
For gamma rays of energy $E_{\gamma}$ the cross section for the $\gamma\gamma$ interaction peaks for low-energy photons with energy:
\begin{equation}
\epsilon\simeq\frac{2m_e^2c^4}{E_{\gamma}}\simeq (0.5 \mbox{ eV})\left(\frac{1\mbox{ TeV}}{E_{\gamma}}\right),
\end{equation}
where $m_e$ is the electron mass and $c$ is the speed of light. 
In terms of wavelength \cite{2011MNRAS.410.2556D}: $$\lambda[\mu\mbox{m}]=1.24E_{\gamma}[\mbox{TeV}].$$ Since HAWC is sensitive between 100 GeV and 100 TeV, observations of AGN will allow us to study the EBL between 0.1 and 100 $\mu$m. This is the range that corresponds mainly to the second peak in the EBL and is dominated by IR emission from starlight absorbed and re-radiated by dust. \\
\section{AGN Sample}\label{sample}
\begin{figure}
\begin{center}
\includegraphics[width=\textwidth ,height=13cm]{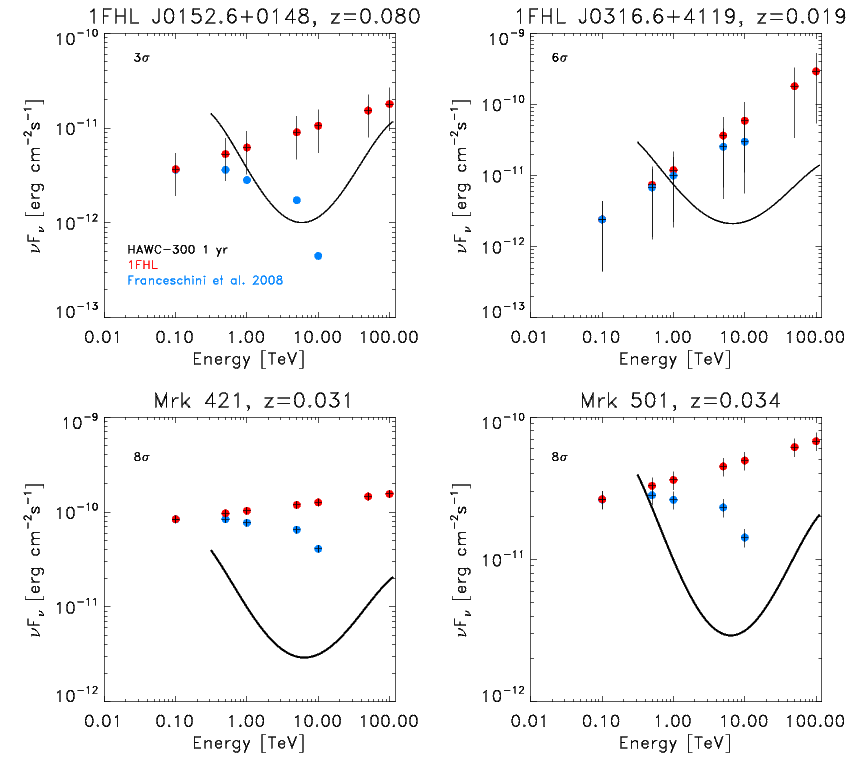}
\caption{Four of 171 1FHL extrapolated spectra to 100 TeV (red dots), the predicted attenuated spectra using and EBL model from \cite{2008A&A...487..837F} (blue dots), and the sensitivity curve for 1 year of observations with HAWC (black line) labeled in the upper left corner. The extrapolation is based on the steady-state spectra from the GeV band to TeV, and we have ignored the possibility of strong flares.}\label{ss1}
\end{center}
\end{figure}

To estimate the sensitivity of HAWC to AGN in the Northern Hemisphere, we selected AGN from the 1FHL catalog, the First Fermi-LAT Catalog of Sources above 10 GeV \cite{2013ApJS..209...34A}. The 1FHL catalog contains 261 AGN in the field of view of HAWC, which decreases to 171 objects when we apply a redshift cut of $z<0.5$.\\
The Large Area Telescope (LAT) aboard the Fermi Gamma-ray Space Telescope detects photons between 20 MeV and 300 GeV. Since the attenuation due to EBL is almost negligible in this energy range, we assume the energy spectra observed with the LAT are close to the intrinsic spectra of the sources \cite{2011ApJ...733...77O}. We extrapolate the fluxes observed at GeV into the TeV band, and then attenuate the spectra using the EBL model of Franceschini et al. \cite{2008A&A...487..837F}. Using the extrapolated and attenuated energy spectra, we can estimate the expected sensitivity of HAWC to the 171 1FHL objects at redshifts below 0.5.\\
Based on the expected sensitivity of the full HAWC detector \cite{hawcsensi}, we can estimate the time needed to detect the extrapolated AGN fluxes at the $5\sigma$ level. Figure \ref{ss1} shows the predicted TeV spectra of four AGN when compared to the 1-year sensitivity curve of the HAWC detector. Using the estimated one-year sensitivity of HAWC and simulated gamma rays from these sources, we found that 12 out of 171 AGN have a high probability of being observed with one year of data. These 12 sources are listed in Table \ref{sample1}. From this list, six sources have variability periods registered with other telescopes. Measuring a spectra during one of the high flux states will provide more statistics in an EBL future work.
\subsection{Markarian 421}
\begin{figure}
\begin{center}
\includegraphics[width=14cm ,height=9cm]{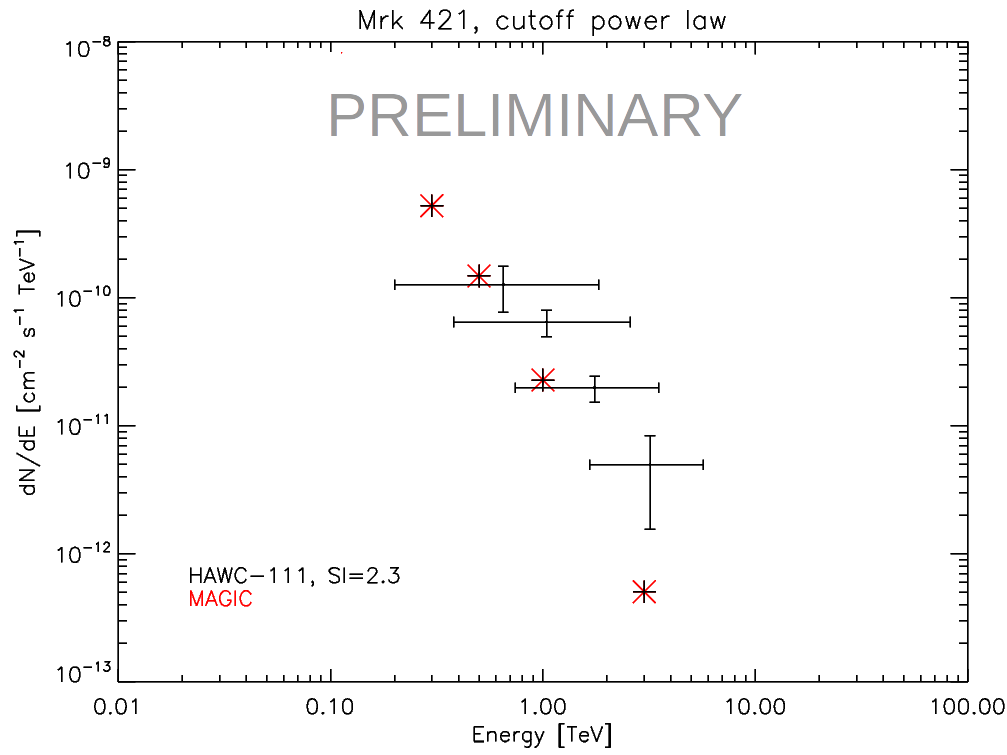}
\caption{Preliminary energy flux measurements of Markarian 421 (black dots) for a cut-off power law in comparison with MAGIC observations \cite{mrk421MAGIC} (red stars). The energy bins are the median energy from the energy distribution in each data bin where the normalization flux was fitted individually in each bin using the likelihood technique. Since there are very large bin-to-bin correlations, it is not a true differential energy spectrum, however we expect to remove the correlations between bins with a more accurate energy proxies.}\label{mrk421DES}
\end{center}
\end{figure}

Markarian 421 (Mrk 421) is one of the most studied gamma-ray AGN as well as the one that shows high degree of variability. It has been observed by HAWC, which found a possible correlation between the TeV and X-ray activity \cite{bib:lightcurves}.The fact that its distance from us is relatively short provides the opportunity to study the EBL using very high energy photons coming from it. \\
As described in \cite{andy}, HAWC data is divided in 10 bins, in each of them the flux normalization was calculated individually using the Likelihood Fitting Framework (LiFF), described in detail in~\cite{bib:liff}. It was found that the normalization flux depends very weakly on the assumed spectral index. Taking the median energy in each bin, preliminary energy flux measurements for a cut-off power law were calculated with 180 days of data during the construction phase of HAWC (2013/08 - 2014/03). Figure \ref{mrk421DES} shows that this preliminary flux measurement is in agreement with MAGIC observations \cite{mrk421MAGIC}. Since there are very large bin-to-bin correlations, it is not a true differential energy spectrum, however we expect to obtain  spectra with more accurate energy proxies to remove the correlations between bins. \\
\begin{table}
\begin{center}
\begin{tabular}{cccccc}
\hline 
1FHL & Alternative name & Type & $z$ & $\Gamma$ & $\sigma/\sqrt{yr}$\tabularnewline
\hline 
J0035.9+5950 & 1ES 0033+595 & bzb & 0.086 & $1.74\pm0.18$ & 6.02\tabularnewline
J0152.6+0148 & PMN J0152+0146 & bzb & 0.080 & $1.77\pm0.34$ & 4.85\tabularnewline
J0316.6+4119 & IC 310 & rdg & 0.019 & $1.31\pm0.45$ & 13.16\tabularnewline
J0521.7+2113 & VER J0521+211 & bzb & 0.118 & $1.97\pm0.14$ & 3.02\tabularnewline
J0650.8+2504 & 1ES 0647+250 & bzb & 0.203 & $1.56\pm0.18$ & 10.25\tabularnewline
J0816.3-1310 & PMN J0816-1311 & bzb & 0.046 & $2.06\pm0.27$ & 3.19\tabularnewline
J1104.4+3812 & Mrk 421 & bzb & 0.031 & $1.91\pm0.06$ & 6.23\tabularnewline
J1230.8+1224 & M 87 & rdg & 0.004 & $1.25\pm0.50$ & 20\tabularnewline
J1653.9+3945 & Mrk 501 & bzb & 0.034 & $1.86\pm0.10$ & 5.30\tabularnewline
J1728.3+5014 & I Zw 187 & bzb & 0.055 & $1.67\pm0.34$ & 3.85\tabularnewline
J2322.5+3436 & TXS 2320+343 & bzb & 0.098 & $1.51\pm0.32$ & 9.68\tabularnewline
J2347.0+5142 & 1ES 2344+514 & bzb & 0.044 & $1.48\pm0.18$ & 5.14\tabularnewline
\hline
\end{tabular}
\caption{Selected sources to study the EBL. The first column is the name in the 1FHL catalog, the second one is the alternative name given in the same catalog; the third column is the type of galaxy (bzb: blazar, rdg: radio galaxy), the fourth column is the redshift of the source, the fifth column is the spectral index in the 10-500 GeV regime given in the catalog, and the last column is the expected significance level of the detection during the first year of observations.}\label{sample1}
\end{center}
\end{table}
\section{Conclusion}
The selected extragalactic sources shown in Table \ref{sample1} are the most likely to be detected within the first year of observations with HAWC, all of them have a redshift below 0.3, which according to the Figure \ref{ze}, will be observed at $E>10$ TeV. And over 180 days of data taken between August 2013 and March 2014 with an early configuration of the HAWC  Observatory have been analysed with a likelihood technique to produce flux measurements of Mrk 421. More days of data are needed to start an EBL formal study but with data taken with only a third of the detector array we have demonstrated the capability to detect Mrk 421 and the opportunity to study the EBL in a wavelength region that has not been constrained yet.
\section*{Acknowledgments}
\footnotesize{
We acknowledge the support from: the US National Science Foundation (NSF); the US Department of Energy Office of High-Energy Physics; the Laboratory Directed Research and Development (LDRD) program of Los Alamos National Laboratory; Consejo Nacional de Ciencia y Tecnolog\'{\i}a (CONACyT), Mexico (grants 260378, 55155, 105666, 122331, 132197, 167281); Red de F\'{\i}sica de Altas Energ\'{\i}as, Mexico; DGAPA-UNAM (grants IG100414-3, IN108713,  IN121309, IN115409, IN113612); VIEP-BUAP (grant 161-EXC-2011); the University of Wisconsin Alumni Research Foundation; the Institute of Geophysics, Planetary Physics, and Signatures at Los Alamos National Laboratory; the Luc Binette Foundation UNAM Postdoctoral Fellowship program.}

\end{document}